\documentclass{article}

\usepackage{arxiv}

\usepackage[utf8]{inputenc} 
\usepackage[T1]{fontenc}    
\usepackage{hyperref}       
\usepackage{url}            
\usepackage{booktabs}       
\usepackage{amsfonts}       
\usepackage{nicefrac}       
\usepackage{microtype}      
\usepackage{graphicx}
\usepackage{lipsum}

\title{In-line Image Transformations for Imbalanced, Multiclass Computer Vision Classification of Lung Chest X-Rays}

\author{
  Alexandrea K. Ramnarine\\
  School of Professional Studies\\
  Northwestern University\\
  Chicago, IL 60611 \\
  \texttt{alexandrearamnarine2021@u.northwestern.edu} \\
}

\begin{document}
\maketitle

\begin{abstract}

Artificial intelligence (AI) is disrupting the medical field as advances in modern technology allow common household computers to learn anatomical and pathological features that distinguish between healthy and disease with the accuracy of highly specialized, trained physicians. Computer vision AI applications use medical imaging, such as lung chest X-rays (LCXRs), to facilitate diagnoses by providing “second-opinions” in addition to a physician’s or radiologist’s interpretation. Considering the advent of the current Coronavirus disease (COVID-19) pandemic, LCXRs may provide rapid insights to indirectly aid in infection containment, however generating a reliably labeled image dataset for a novel disease is not an easy feat, nor is it of highest priority when combating a global pandemic. Deep learning techniques such as convolutional neural networks (CNNs) are able to select features that distinguish between healthy and disease states for other lung pathologies; this study aims to leverage that body of literature in order to apply image transformations that would serve to balance the lack of COVID-19 LCXR data. Furthermore, this study utilizes a simple CNN architecture for high-performance multiclass LCXR classification at 94\% accuracy. \footnote{Code available at \url{https://github.com/ramnadime/COVID19_LCXR_multiclassification}}
\end{abstract}


\section{Introduction}
\label{sec:Intro}
Lung diseases are often misdiagnosed based on medical imaging alone or require expensive and non-point-of-care diagnostics, such as in the case of COVID-19. Medical imaging can be expensive, is usually time-consuming, and exposes patients to risks such as radiation that may compromise health later in life with frequent exposure. Furthermore, analyzing medical images must be performed by trained, specialized physicians, including pathologists and radiologists, which is an arduous and time-consuming task given large amounts of images, and ultimately relies on physician expertise to minimize subjectivity as much as possible. Specific to X-ray imaging, LCXRs pose many problems, such as common diagnosis confusion with Pneumonia, bone and device occlusion, and high-noise low-contrast imaging. Implementation of artificial intelligence, specifically deep computer vision applications, can increase diagnostic accuracy through feature learning on computer-aided enhanced images, which would speed up patient triage and treatment processes.

\section{Literature Review}
\label{sec:LitRev}
\setlength{\parindent}{1em}
Within the last decade, a few companies successfully leveraged artificial intelligence to change the way physicians approach medical diagnoses, such as \cite{contextvision_2018} deep learning platform for radiography image enhancement. That companies can secure funding to develop such technologies demonstrates the acknowledgment of computer vision as a potentially indispensable medical tool; however, given the sensitivity of the use case, much work needs to be done to ensure that the outputs of these computation tools do not compromise human lives. Current reviews of medical computer vision warn of uninterpretable “black-box” algorithms dictating diagnoses \cite{hosny_parmar_quackenbush_schwartz_aerts_2018}. While it is difficult to determine what deep networks are learning on without parsing through every activated node in a network, CNNs are particularly easy to take apart; each activated filter can be visualized to understand the exact pixel features a model identified per image, essentially negating the opinion that computer vision models are ambiguous in nature.

\cite{SHIRAISHI2011449} report artificial neural network performances in a range of 0.9 to 0.97 area under the curve values on lung images, indicating that computer vision models can achieve very high classification rates on complex, multiclass image sets. Given the reliable accuracies deep learning can achieve, the study suggests that these models do not have to be very complex in architecture. Rather, changes to the images themselves using filters to manipulate pixel representations can be utilized to fine-tune model performance, such as accuracy and loss.	

Chest imaging is commonly used to support medical diagnoses, particularly X-rays, which are affordable and easy to procure \cite{qin_yao_shi_song_2018}. Qin et al.’s survey of AI preprocessing of LCXRs mention image enhancements, such as contrast and edge enhancement techniques, and image segmentation and bone suppression to better focus in on regions of interest. This study focuses on the former, utilizing contrast and edge enhancement methods developed for general graphic and photography manipulation. The idea is to use the simplest techniques to transform the data as few times as possible, without having to remove or crop images, in order to still achieve high classification rates. COVID-19 LCXRs are new as of 2020 with the onset of the pandemic, thus open-sourced, reliably labeled image data are short in supply. To increase class representation using reliably sourced COVID-19 images, image transformations such as flips and rotations can be utilized to provide additional training data for the models in a manner still consistent with clinical relevancy \cite{Goldstein2020.10.01.20204073}.

\cite{math8040545} implement two mathematical functions to perform only contrast enhancement, histogram equalization and gamma correction. These methods ignore the strengths of edge detection filters commonly used in graphic design. While this study explores simpler contrast enhancement techniques, as well as a wider gamut of transformations, it follows Chen et al.’s suggestion of implementing fully convolutional networks to satisfy the classification task without fully-connected layers within the architecture. To try an alternative schema, a dense layer is implemented after all convolutions in this study to assess if it can achieve greater than 90\% accuracy without utilizing image segmentation techniques as Chen et al. do in their study. 

Model parameters are mainly dictated by CNN architecture. Fully convolutional networks for medical imaging regarded as “complex” can have 150,000 parameters training on images the size of 256x 256 \cite{shin_roth_gao_lu_xu_nogues_yao_mollura_summers_2016}. Shin et al. compare three CNN architectures with two or more convolutions in each to achieve a maximum of 95\% classification accuracy. The CNN architecture used in the study herein has 107,332 parameters contributed by only two convolutional layers for 100x100 pixel resized images, however utilizing a dense layer adds 4,956,320 additional model parameters; additional experiments can assess the performance of these models without the additional dense hidden layer. It should be noted that Shin et al. utilize pre-trained models on medically irrelevant data and transferred learning layers, which is a potential continuation but not a main focus of this current study.

In a recent clinically relevant study also from Northwestern, \cite{doi:10.1148/radiol.2020203511} used an ensemble of popular, pretrained CNNs on pre-processed LCXRs to achieve about 82\% test accuracy for binary classification of COVID status. This ensemble deep learner, entitled “DeepCOVID-XR”, performed on par with the consensus labeling of five thoracic radiologists. The radiologists examined a random set of 300 images on the order of hours, while DeepCOVID-XR was able to scan the entire set in only 18 minutes.

\section{Methods}
\label{sec:Methods}

\subsection{Data Sources}
The LCXR data downloaded for use in this study were organized and collected by Kaggle user Prashant Patel from the following sources: Cohen et al.’s COVID-19 image dataset \cite{cohen2020covid}, Kermany et al.’s image-based deep learning dataset \cite{kermany_goldbaum_cai_valentim_liang_baxter_mckeown_yang_wu_yan_2018}, and Chung et al.’s COVID-Net team \cite{wang_wong_qiulin_mcinnis_chung_gunraj_lee_ross_vanberlo_ebadi_2020}. The LXCRs fall into three diagnosis classes: Normal, COVID-19, and non-COVID-19 Pneumonia.

\subsubsection{Correcting Dataset Imbalance}
Appendix A demonstrates the class imbalance between Normal, COVID-19, and Pneumonia LCXR images. A CNN trained on the dataset as is could only predict class 2 (Pneumonia) for the test set (not shown), thus a correction to the class representation was performed. To correct imbalance, no images were added to either training or test sets for Pneumonia images, since this class had the largest representation from the Original set. Training images were handled in separate directories from test images to avoid data leakage. In Windows Picture Tools, all training and test class 0 (Normal) photos were rotated 180 degrees in Windows Picture Tools and added to respective directories. All training and test class 1 (COVID-19) photos were rotated 90 degrees to the left, then Original images were rotated 90 degrees to the right, then Original images were rotated 180 degrees in Windows Picture Tools and added to directories. The resulting numbers of images per class are also found in Appendix A, leading to a more balanced complete dataset.

\subsubsection{Image Transformation}
All images were read into a Jupyter notebook running on Python 3.7.9 using Pillow, a fork for PIL, and glob packages. Training and test images were stored locally in separate directories, and thus were loaded into Python separately for separate image normalization and transformation to avoid data leakage. Images were converted to 100x100 pixels, converted to arrays, and normalized by dividing by 255 to convert the Original dataset to a NumPy array. In order to apply image filters to the Original dataset, the dataset was read in four additional separate times to create four transformed datasets using PIL image filters “CONTOUR”, “EDGE ENHANCE MORE”, “FIND EDGES”, and “SHARPEN”. After transformation, the respective images were converted to 100x100 pixels, normalized and converted to arrays in the same manner as the Original dataset.

\subsubsection{Hyperparameter Optimization}
A simple CNN architecture was chosen to model all datasets in TensorFlow 2.0 using the high-level Keras API. A Sequential model consisted of the following layers in this order: 2-dimensional convolution (Conv2D), 2-dimensional max pooling (MaxPool2D), Conv2D, MaxPool2D, flatten, dense, dense output. Models utilizing dropout regularization had two dropout layers, one after each MaxPool2D layer. All MaxPool2D layers used a 2x2 pool size. All Conv2D layers utilized a He uniform kernel initializer for weight assignment. Each Conv2D and dense layers utilized rectified linear units activation, and the final dense output layer utilized softmax activation. The Adam optimizer and sparse categorical cross entropy loss function were used for model compilation. Keras Tuner implementation of the Hyperband tuner, with a validation accuracy objective, maximum epoch of 15, and factor of 3 was used to find an optimal value for number of Conv2D layer filters, Conv2D kernel size, number of dense layer units, and Adam optimizer learning rate using the Original dataset. Tuner returned an optimization of 160 dense units, 64 convolutional filters, a 5x5 kernel size, and a 0.0001 learning rate. These hyperparameters were passed to each model architecture, and the resulting model from the Hyperband search was used to train on the Original dataset.

\subsubsection{CNN Training}
Each dataset was trained with the same CNN model architecture and hyperparameters to control only for dataset variations by the image filter transformations. 30\% of the training data were shuffled for each of the 15 epochs to use as a validation set. Training times were recorded, and accuracy and loss were collected for training, validation, and test sets. Scikit-Learn implementations of confusion matrix and classification report were utilized for each model to understand class-stratified metrics. Seaborn was used to visualize the confusion matrices and MatPlotLib was utilized to generate accuracy and loss plots, as well as visualize the dataset images themselves. Model summaries were generated, and the trained models were saved as TensorFlow models for future use. 

\subsubsection{CNN Filter Extraction}
One randomly selected test image was visualized using MatPlotLib and converted to an image tensor for CNN layer filter extractions. Output of all CNN layers per model were extracted, and a model was created for each dataset instance (ie, each model ran during training) that returned the activation outputs. These activations were used to predict on the image tensor and then displayed as a feature map using MatPlotlib.

\section{Results}
\label{sec:Results}

\subsection{Image Transformations}
The image filters created a wide variety of pixel transformations on the Original dataset. Figure 1 shows randomly selected training images prior to class balance with filters applied. The Contoured and Found Edge datasets are visually the most dramatic transformations for the LCXR images with almost complete color inversions and feature reduction. The More Enhanced Edge and Sharpened datasets are visually closer to the Original dataset, but it is apparent in both cases that features appear more pronounced relative to the Original set. After balancing the dataset, all COVID-19 and Normal labeled images were rotationally transformed. A subset of these training images, which were used for final model building, are shown in Figure 2; images in the second and fifth column for all five datasets are rotated 180 and 90 degrees to the left, respectively. After rotational transformation, these images were then subjected to image filter transformation. Given the further reduction in image size for model training, the transformations are even more pronounced. More features are visually apparent in the Contoured and Found Edge datasets, and the More Enhanced Edge dataset is now obviously visually different from the Original dataset. 

\begin{figure}[h]
\caption{Select Training Images from the Original Dataset Prior to Class Balancing}
\includegraphics[width=\textwidth]{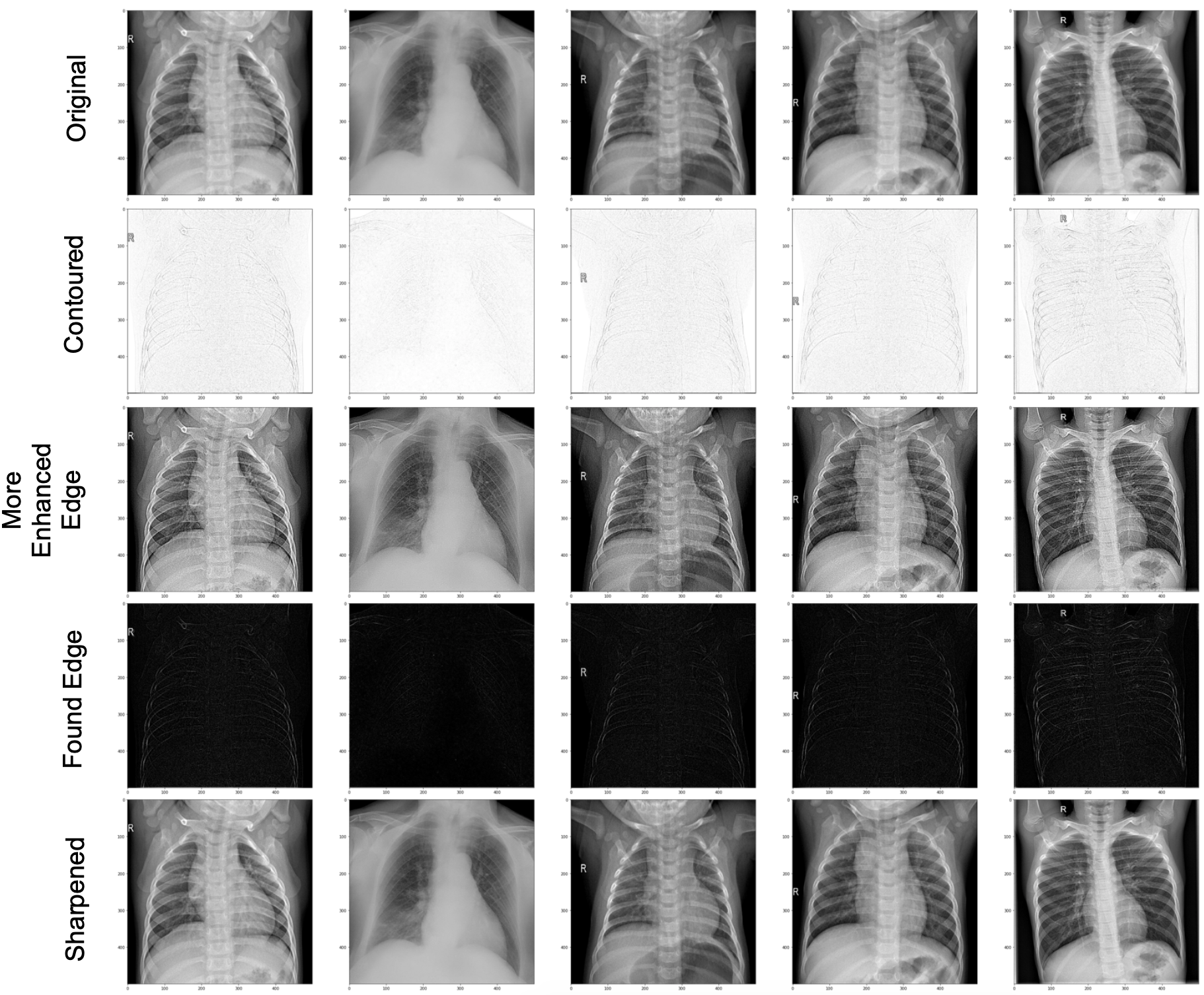}
\emph{Note:} These images were resized to 500x500 and were not used for modeling.
\end{figure}

\begin{figure}[h]
\caption{Select Training Images from the Original Dataset After Class Balancing}
\includegraphics[width=\textwidth]{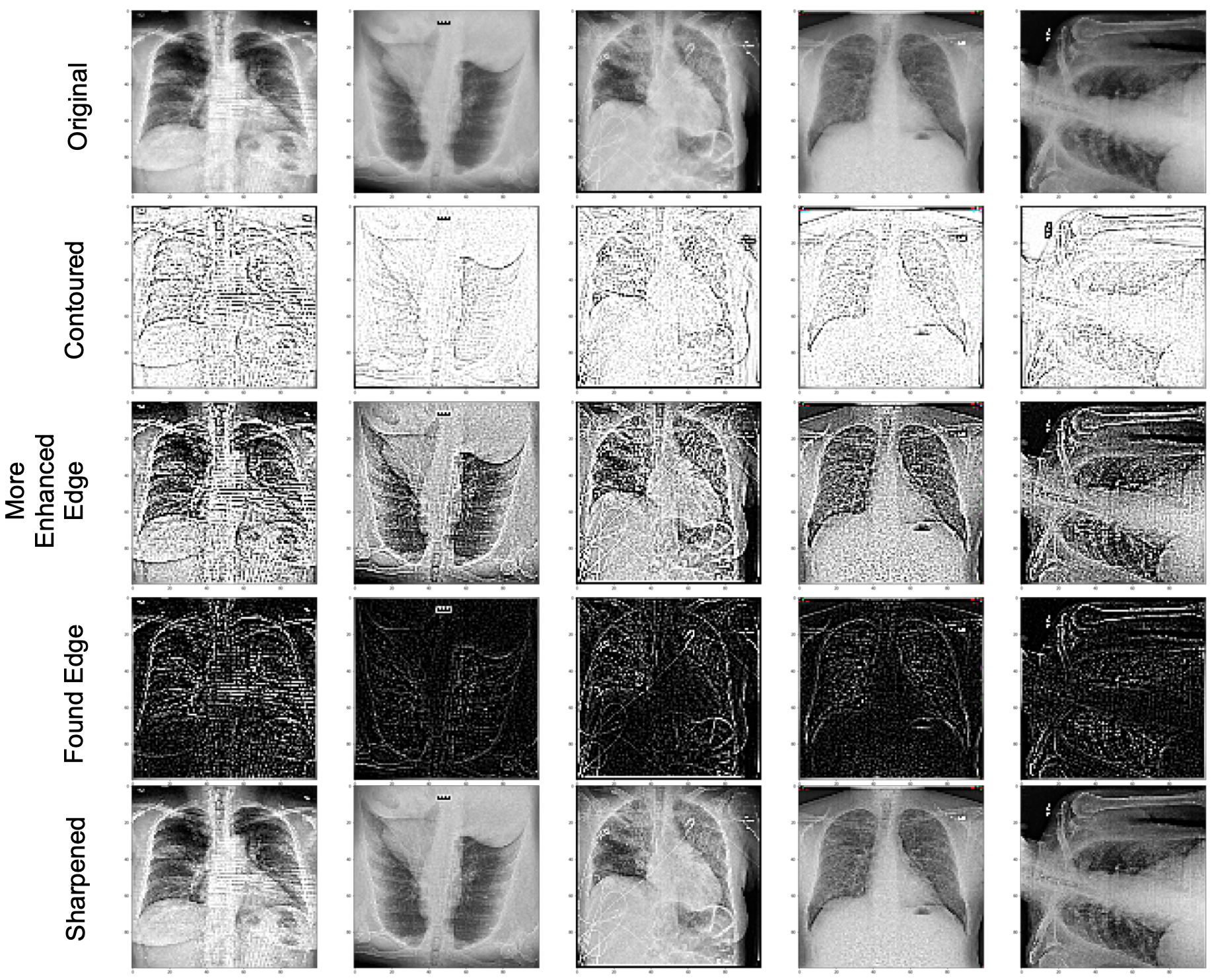}
\emph{Note:} These images are resized to 100x100 and were used for modeling.
\end{figure}

\subsection{CNN Performance}
Seven different CNN models were trained using the Original and transformed datasets, where the last two models correspond to a CNN utilizing dropout regularization to train on the best performing dataset. Table 1 shows the processing times and training, validation, and test metrics for each model based on the dataset transformation used. Models are here on referred to by the names of the dataset they trained on.

\begin{table}[h!]
 \caption{Comparison of Training, Validation, and Test Metrics for All Models}
  \centering
  \begin{tabular}{lccccccc}
    \toprule
    \multicolumn{2}{c}{}&\multicolumn{2}{c}{Training}&\multicolumn{2}{c}{Validation}&\multicolumn{2}{c}{Test}                   \\
    \cmidrule(r){3-8}
    Transformation     & Process Time (s)     & Acc.     & Loss     & Acc.     & Loss     & Acc.     & Loss \\
    \midrule
    Original & 30.2	& 98.37	& 0.563	& 90.80	& 0.642	& 94.42	& 0.609 \\
    Contour & 30.2 & 99.21 & 0.008 & 86.99 & 0.476 & 91.81 & 0.334 \\
    More Enhanced Edge & 30.4 & 100.0 & 0.001 & 91.10 & 0.403 & 93.65 & 0.264 \\
    Found Edge & 29.9 & 100.0 & 0.001 & 87.38 & 0.556 & 92.17 & 0.321 \\
    Sharpen & 31.0 & 100.0 & 0.003 & 92.30 & 0.319 & 94.21 & 0.235 \\
    Sharpen, 20\% Dropout & 32.3 & 98.92 & 0.030 & 89.43 & 0.306 & 92.32 & 0.221 \\
    Sharpen, 50\% Dropout & 33.1 & 97.56 & 0.070 & 89.43 & 0.282 & 91.86 & 0.226 \\
    \bottomrule
    \emph{Note:} Acc. = Accuracy
  \end{tabular}
  \label{tab:table}
\end{table}

The average processing time for all models was 30 seconds, where the dropout regularization implementation increased training time. In terms of test metrics, the Sharpened dataset without dropout regularization performed best, with a 0.21\% reduction in test accuracy but almost one-third of the test loss relative to the model trained on the Original dataset. Dropout regularization was thus applied to a model trained on the Sharpened dataset, where no improvements in test accuracy were observed, but both dropout percentages favorably decreased test loss, with 20\% dropout decreasing loss the most. 

Relative to the Original dataset, applying transformations drastically decreased test loss in every instance. Contoured and Found Edge transformations halved the loss whereas More Edge Enhanced and Sharpened transformations nearly reduced test loss to a third of the Original test loss. While none of these models outperformed the Original in test accuracy, Contoured and Sharpened models had higher validation accuracies. The test accuracies for the transformed dataset models only varied from the Original model within a range of 0.21\% to 2.61\%.

\subsubsection{Training Metrics}
Plots of training and validation metrics, accuracy and loss, over all 15 training epochs are visualized in Appendix B for all models. In every dataset case, training accuracy increases greatly from epoch 1 to 2, then gradually increases in small increments over the remaining epochs. A similar trend is observed for training loss, where loss decreases rather than increases over epochs. More Enhanced Edge and Found Edge models plateau for both training accuracy and loss around epoch 9. The validation trends for the Original data are spurious across epochs compared to the models trained on transformed images. The Contoured validation trend over all 15 epochs is closest to its respective training trend for both accuracy and loss, followed by the More Enhanced Edge validation trend. The Found Edge validation loss increases over epochs, a trend not observed for any of the other models. Utilizing dropout regularization decreases the spurious validation loss trend observed with the Sharpened model.

\subsection{Classification}
Appendix C provides the confusion matrices for every test output. For normal LCXR images, only the More Edge Enhanced and Sharpened models misclassified two images as COVID-19, the rest had no misclassifications. For the Sharpened model in this instance, dropout regularization reduced those two misclassifications to zero. The Pneumonia class had the largest misclassification rate (~10\%) for every model instance, where it was falsely labeled as Normal; dropout regularization did not lower the rate in the case of the Sharpened model. Overall, the Sharpened model correctly classified the most Pneumonia LCXRs, The Sharpened model with 50\% dropout correctly classified the most Normal LCXRs, and the Contoured model correctly classified the most COVID-19 LCXRs.

\subsection{CNN Filter Feature Extraction}

\begin{figure}[h]
\caption{Selected Image and Subsequent Transformations for Feature Extraction}
\includegraphics[width=\textwidth]{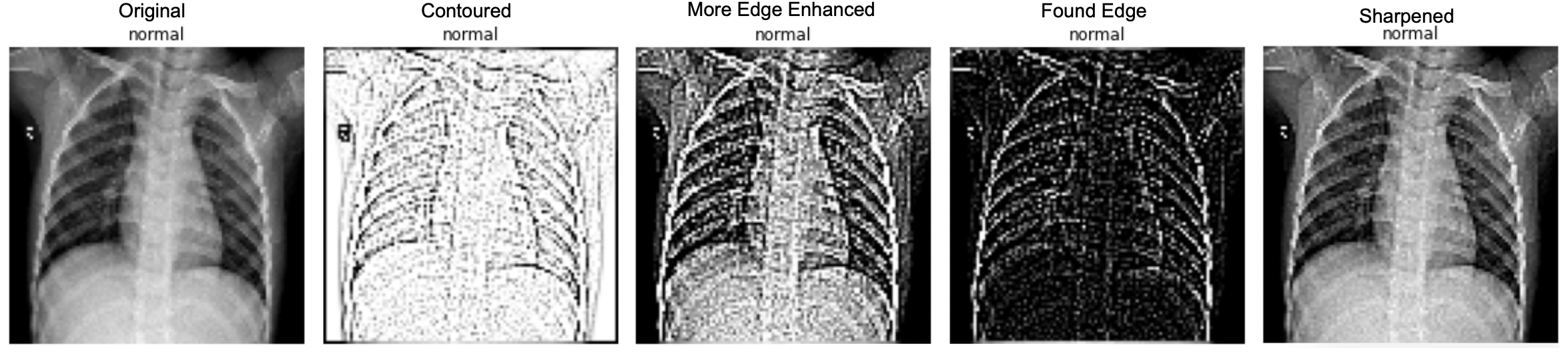}
\end{figure}

\begin{figure}[h]
\caption{First Convolutional Layer Filters of Normal Class LCXR}
\includegraphics[width=\textwidth]{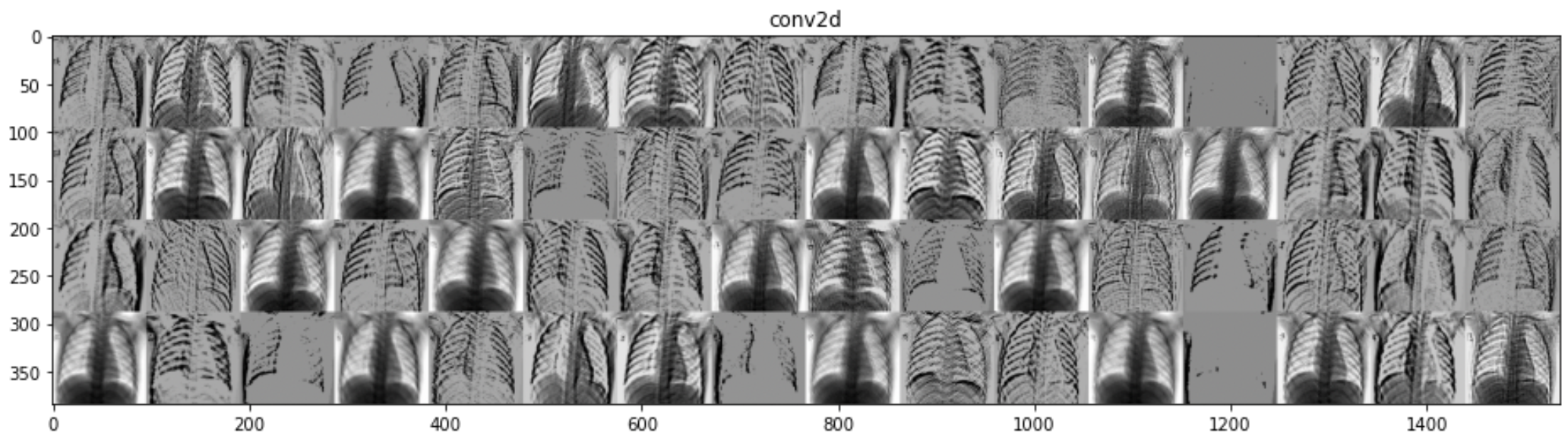}
\end{figure}

Appendices D through J show convolutional and max pooling feature extractions per filter for the same image isolated from each dataset. 64 filters were investigated per convolutional layer with corresponding 64 max pooling feature reductions per filter. The image isolated from the dataset is a Normal class LXCR, shown as Original and with the four different transformations applied separately in Figure 3. The Original, More Edge Enhanced, and Sharpened filters preserve the pixels that correspond to spatial representation of the lung lobes, while Contoured and Found Edge create all white or all black pixel backgrounds respectively.

The first convolutional layers of the Original, More Enhanced Edge, and all Sharpened models have filters that recognized the negative space in the center of the LCXRs, as indicated by the dark pixel regions. Neither of the convolutional layers for the Original model inactivated any filters. A visual example of the extracted first convolutional layer filters from the Normal class image is shown in Figure 4. The first convolutional layer of the Contoured model inactivated one filter, and the second layer inactivated 11 filters. The rest of the transformed image models inactivated filters only in the second convolutional layer: More Enhanced Edge inactivated 34 filters, Found Edge inactivated 41, Sharpened inactivated 36, Sharpened with 20\% dropout inactivated 47, and Sharpened with 50\% dropout inactivated 33.

\section{Discussion}
\label{sec:Discussion}

Using image filters led to more reliable validation trends over training epochs for all models relative to not filtering any of the images. This implies that it is safer to employ early stopping in these filtered instances without worrying about compromising the weight updating of a model by prematurely stopping the training or incorrectly stopping at a poor-performing epoch. In fact, if one were to draw least squares lines for each validation plot in Appendix B, those regression lines would be closer to the training accuracy and training loss values per each epoch for the transformed datasets, implying that those models may generalize better than the model trained on the Original dataset. Future work should include an experiment with dropout regularization per model as dropout seemed to correct spurious validation loss behavior across epochs, which potentially contributed to better class discrimination.

It is difficult to interpret the convolutional feature extractions per filter when only looking at one image per dataset. For this one image, pixels corresponding to ribs seem to still be a main focus for all of the filters. The first filter in the second convolutional layer of the Contoured model appears to focus on the pixel region corresponding to outside of the lung space, which may indicate that cropping, image segmentation and masking may prove useful for these models to focus on the pixels that only correlate to the lungs. That the models trained on transformed images have about half or more inactivated filters indicates that the CNN architecture is deep enough to deduce the encoded patterns within the images. The blank filters indicate that the pattern encoded by that filter, its recognition ability, is not found in this given input image. Future work will include experimenting with a wider array of test images to see if this trend holds true for both Original and transformed datasets, as well as generate heatmaps per image to further elucidate specific pixels that correspond to important model features.

All models discriminated between COVID-19 and Normal LCXRs exceptionally well, but the Sharpened and Contoured models outperformed the Original for this classification. Given the current state of the COVID-19 pandemic, it may prove useful to implement these two image filters for any diagnostic tests using LCXRs using a similar CNN architecture as the one explored in this study. As previous literature alludes to, LCXRs of Pneumonia are misclassified as Normal, and vice versa; in this study however, the Pneumonia images have a much lower misclassification rate with the COVID-19 images. Using image filters may help discern between disease states when computer vision models fail to discern between healthy versus disease, which is a very interesting and unexpected observation from this study. This implies that the vision models are able to learn features correlated to lung pathology, which if true, would refute that computer vision for LCXRs fail due to device and bone occlusions. As more LCXR data becomes available, it would be possible to convert the problem from multiclass to multilabel, as lung pathologies commonly co-present over various disease states. 

There is no clinical or diagnostic value in this study, but it presents an alternative approach to handling medical imaging, particularly LCXRs, in a different manner by applying very basic image filters and transformations used commonly for graphic design and photography editing purposes. There are several impressive feats accomplished by using these drastic image transformations. It was assumed prior to training that dramatic image size reduction, thus subsequent reduction of trainable pixel features, may severely bottleneck model accuracy. Training on image sizes reduced ten-times did not drop test accuracy below 90\%, which is reassuring since LCXRs are typically large file sizes that quickly take up a lot of memory. This indicates that CPUs may be able to handle these models, given the obvious performance time trade-off relative to GPU training, however it increases accessibility and usability for these types of computer vision applications. 

The models trained on the transformed data not only maintain essentially the same test performance, they reduce test loss by 50\% or greater and tame validation metric trends across all epochs relative to the Original model, all without the use of dropout regularization and utilizing a very simple CNN architecture. The filters used in this study are alternatives to very complicated mathematical pixel transformations and image segmentations that previous groups report, making the approach accessible and understandable to a wider range of stakeholders, such as radiologists, physicians, and data scientists alike.

\bibliographystyle{unsrt}  
\bibliography{references}  






\end{document}


\maketitle
Full dataset, code and trained TensorFlow models available at \url{https://github.com/ramnadime/COVID19_LCXR_multiclassification}

\newpage
\section{Appendix A}
\label{sec:A}
\emph{Class Distribution Comparison between Imbalanced vs. Balanced Datasets}
\begin{figure}[ht!]
\includegraphics[width=\textwidth]{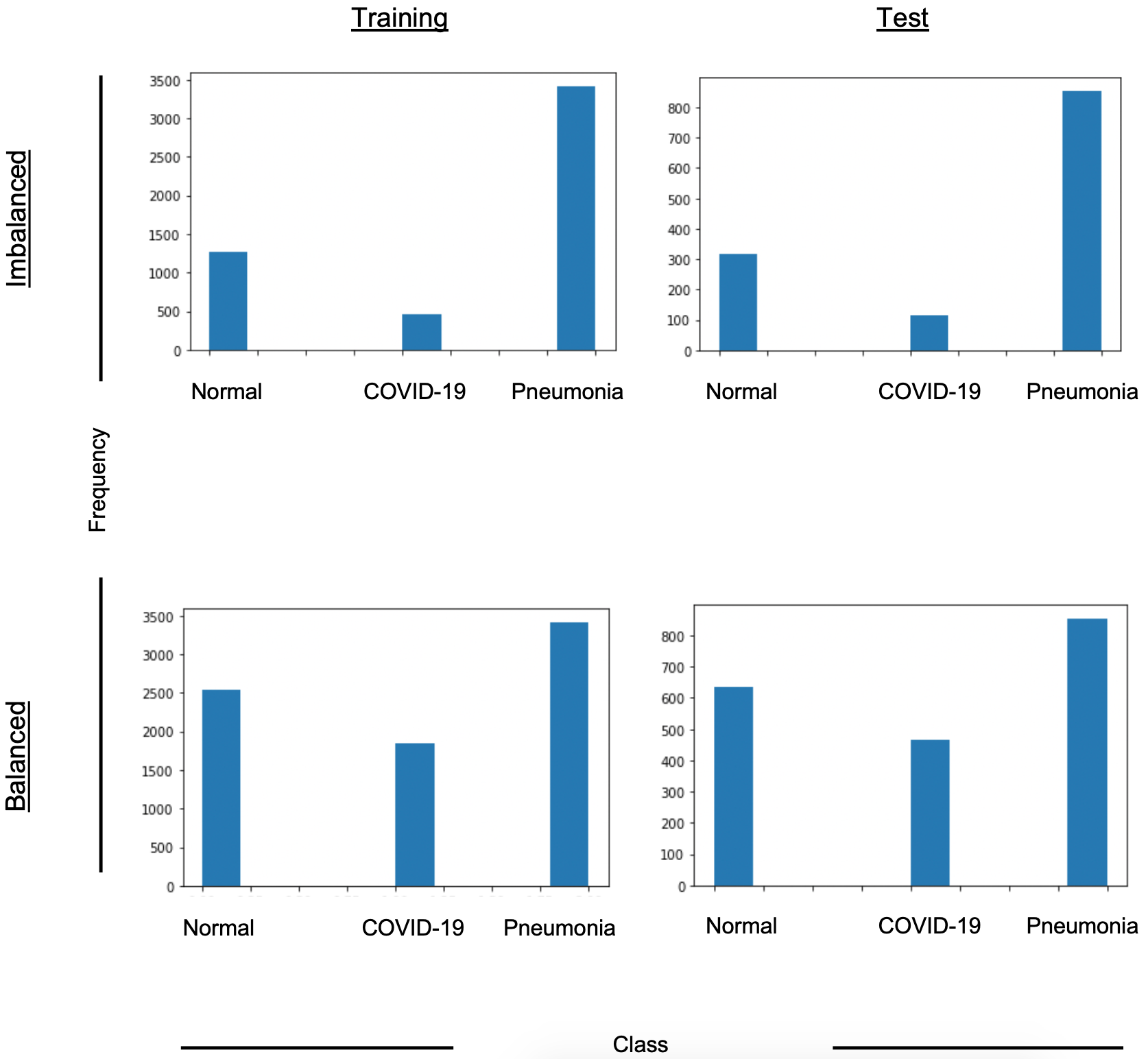}
\end{figure}

\clearpage
\newpage
\section{Appendix B}
\label{sec:B}
\emph{Training and Validation Metrics for All Models (pink validation, blue training)}
\begin{figure}[ht!]
\includegraphics[width=\textwidth]{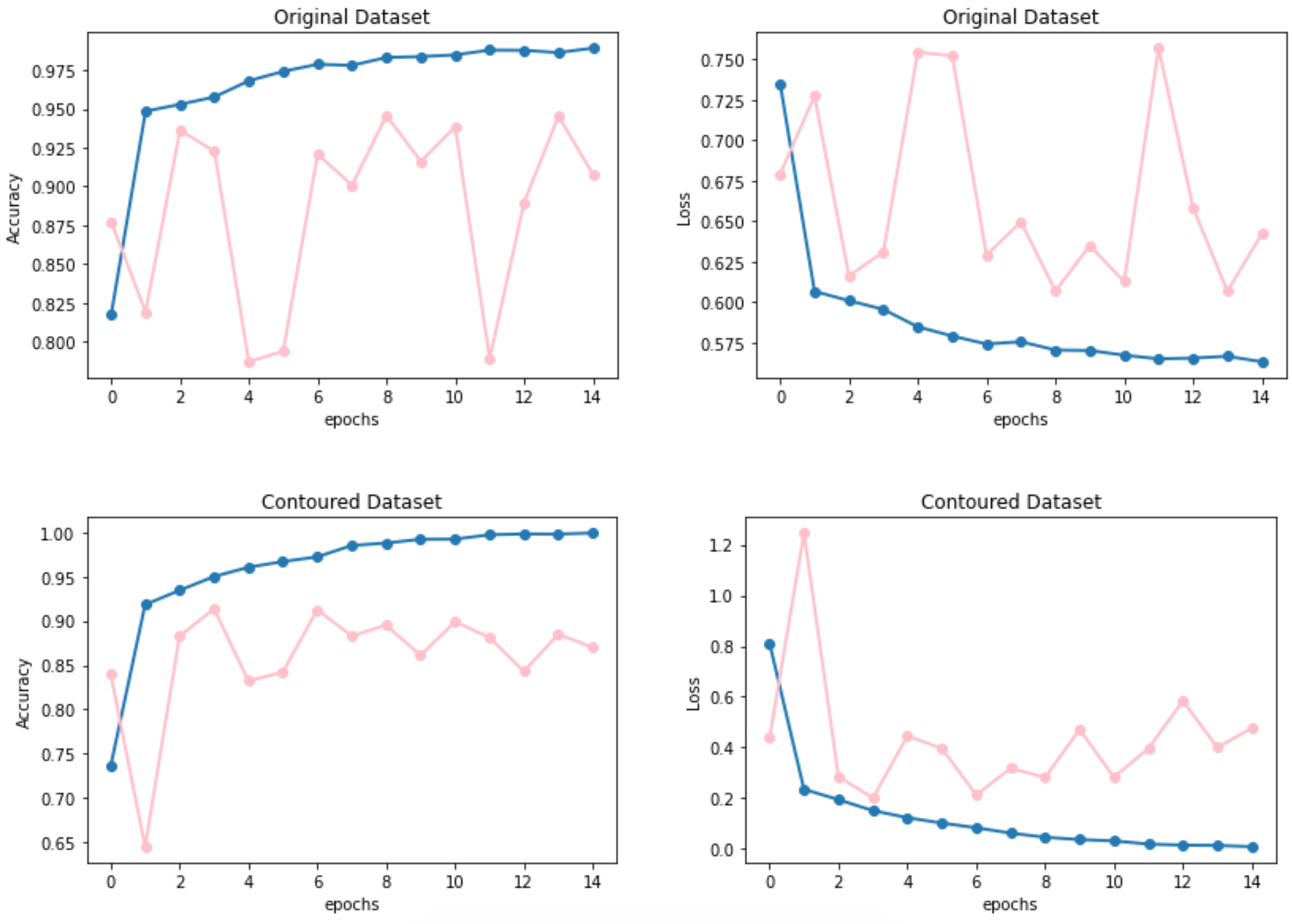}
\end{figure}
\begin{figure}[ht!]
\includegraphics[width=\textwidth]{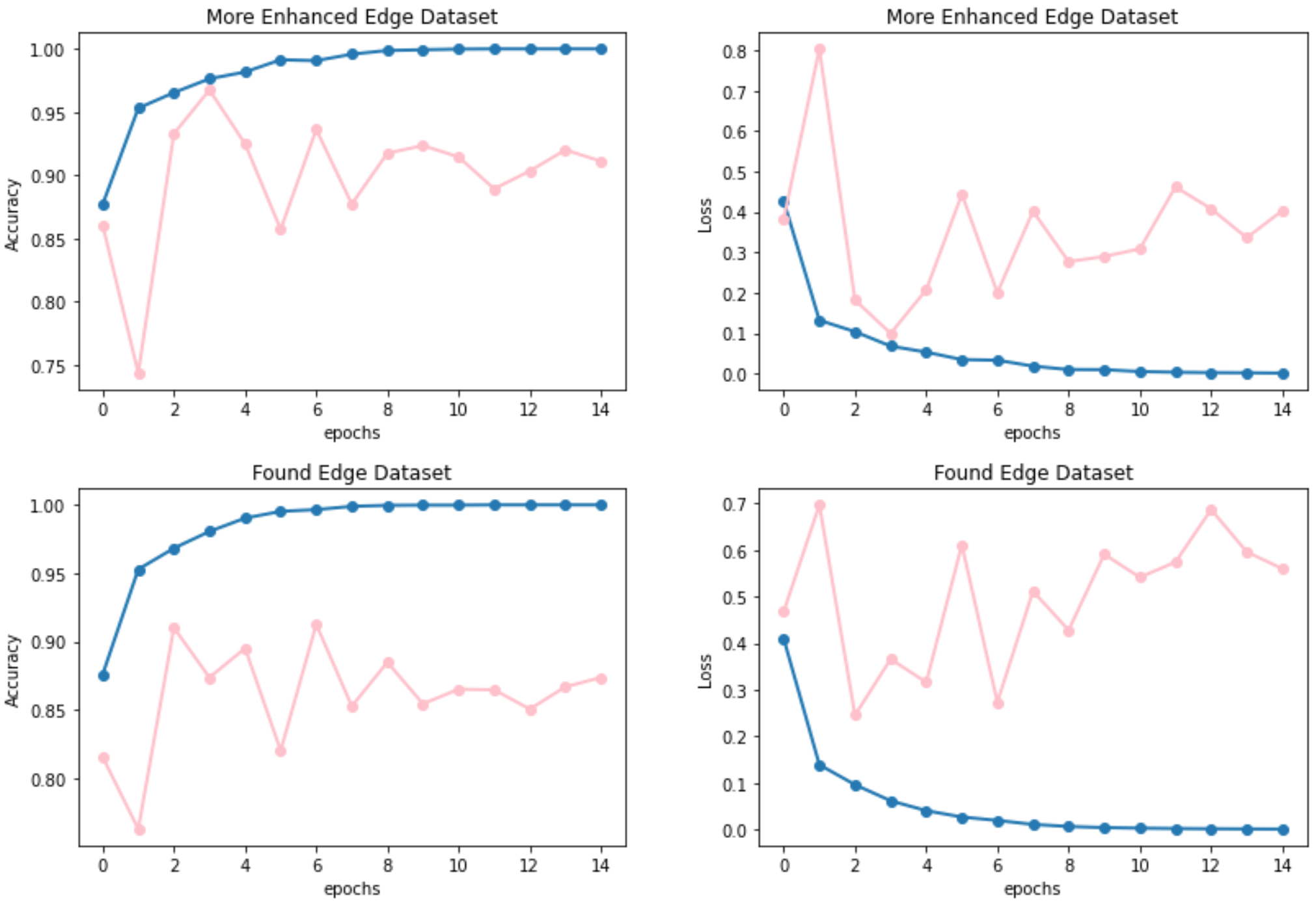}
\end{figure}
\begin{figure}[ht!]
\includegraphics[width=\textwidth]{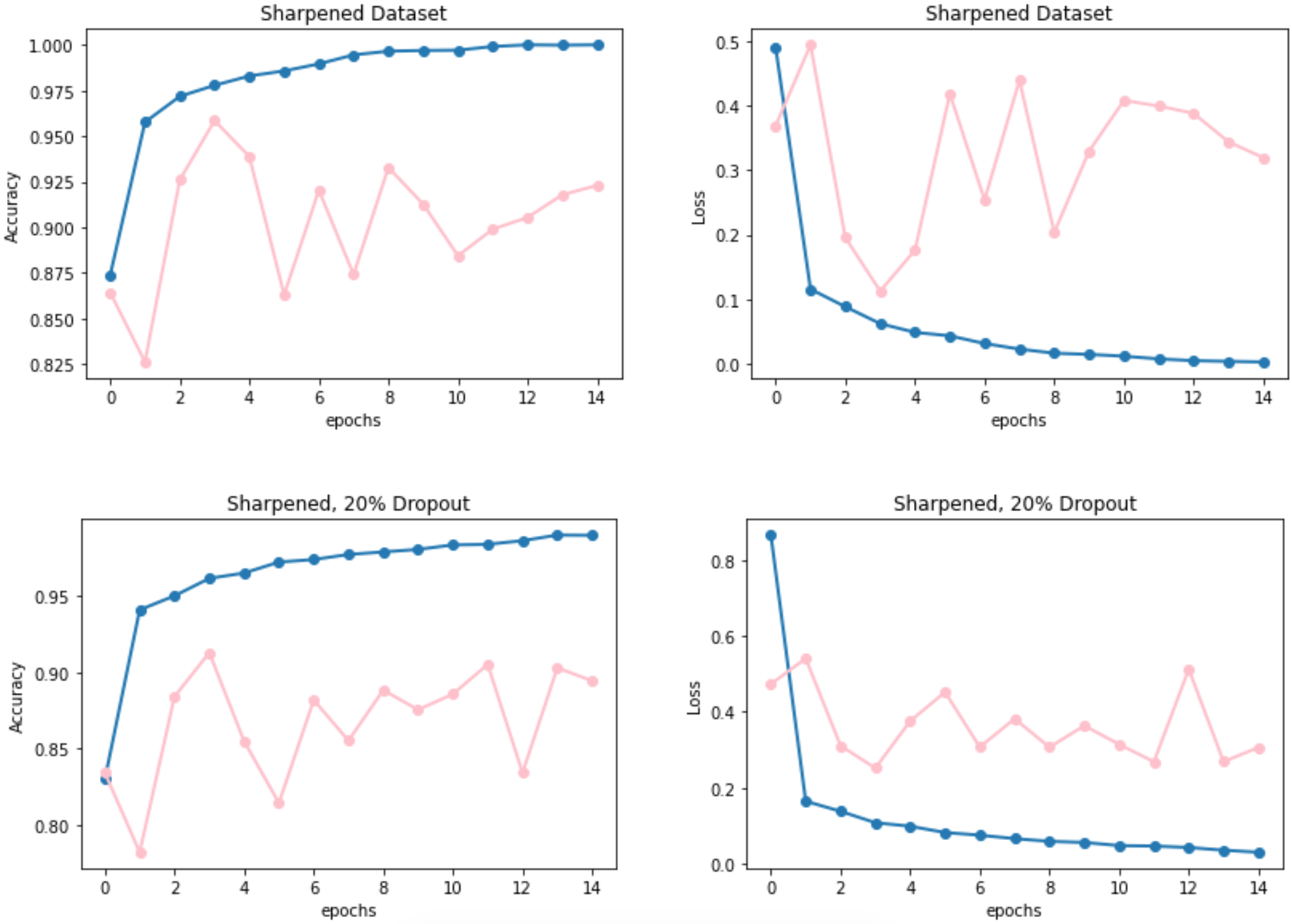}
\end{figure}
\begin{figure}[ht!]
\includegraphics[width=\textwidth]{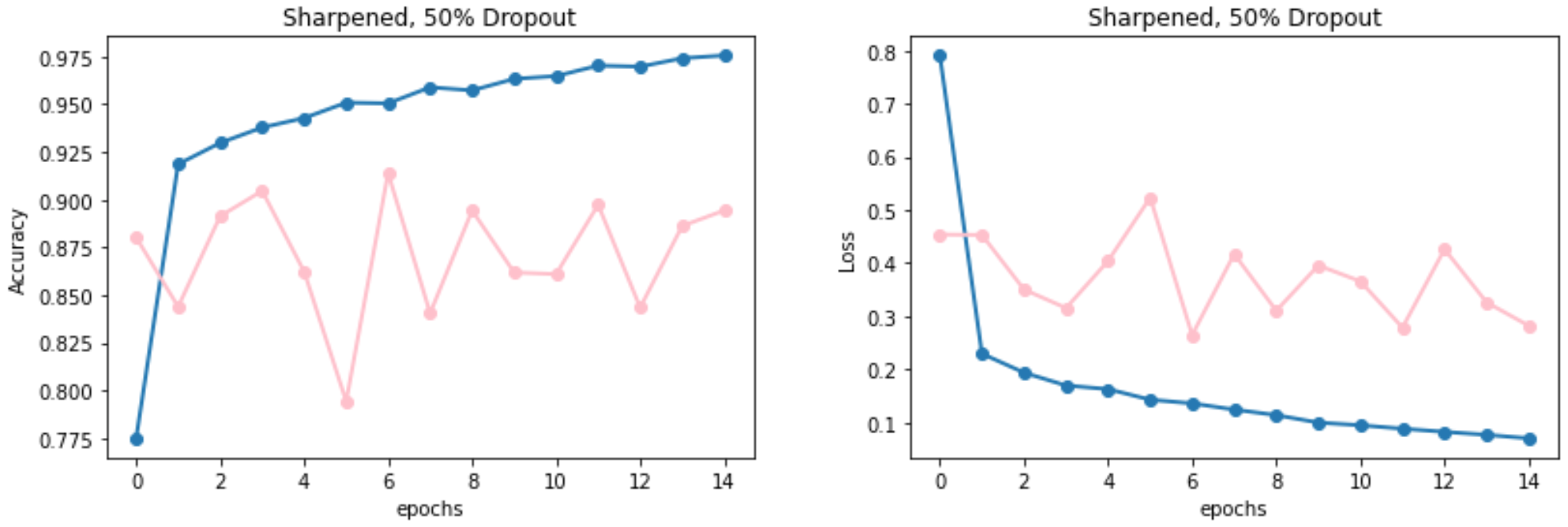}
\end{figure}

\clearpage
\newpage
\section{Appendix C}
\label{sec:C}
\emph{Confusion Matrices for all Models}
\begin{figure}[ht!]
\includegraphics[width=\textwidth]{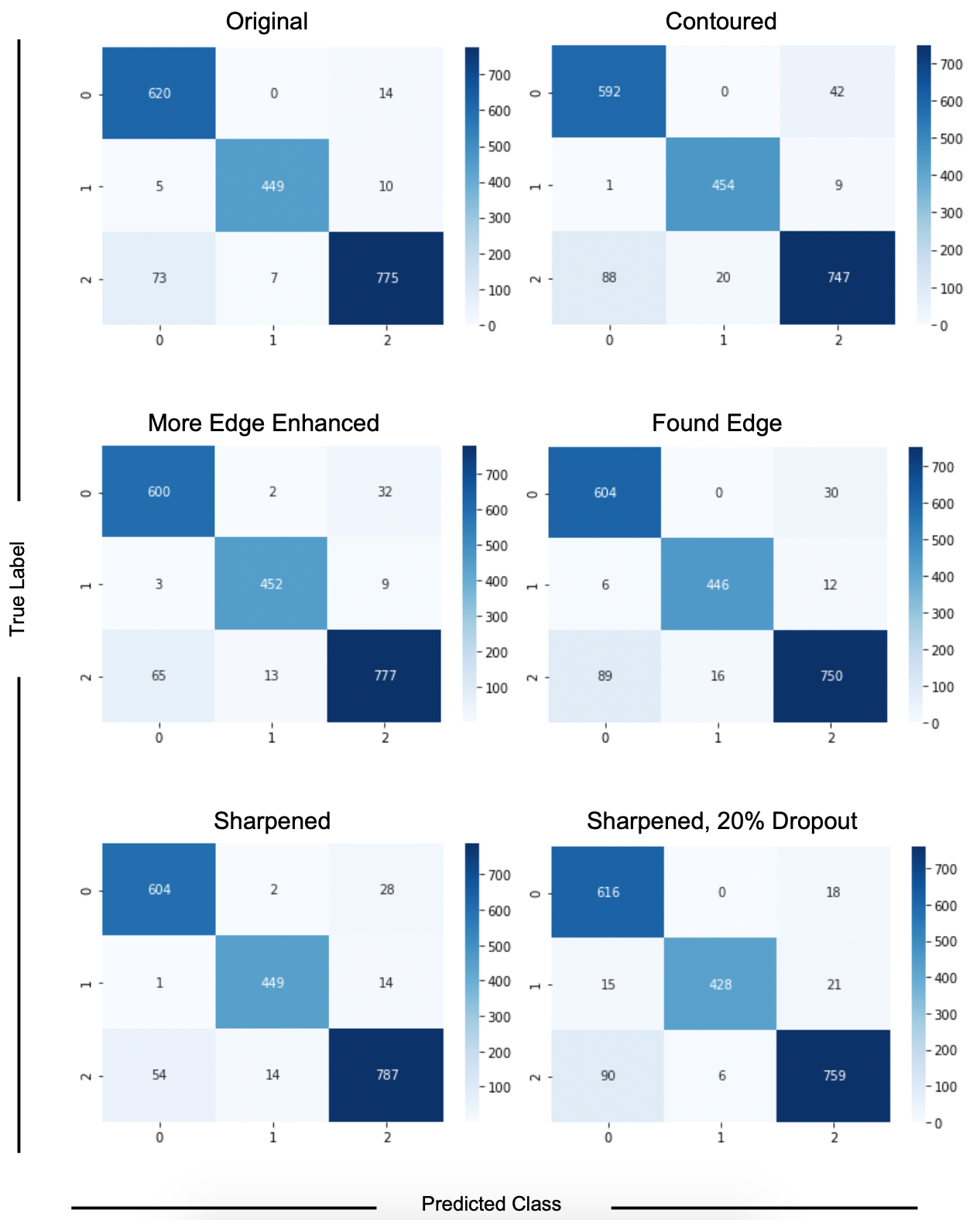}
\end{figure}
\begin{figure}[ht!]
\includegraphics[width=\textwidth]{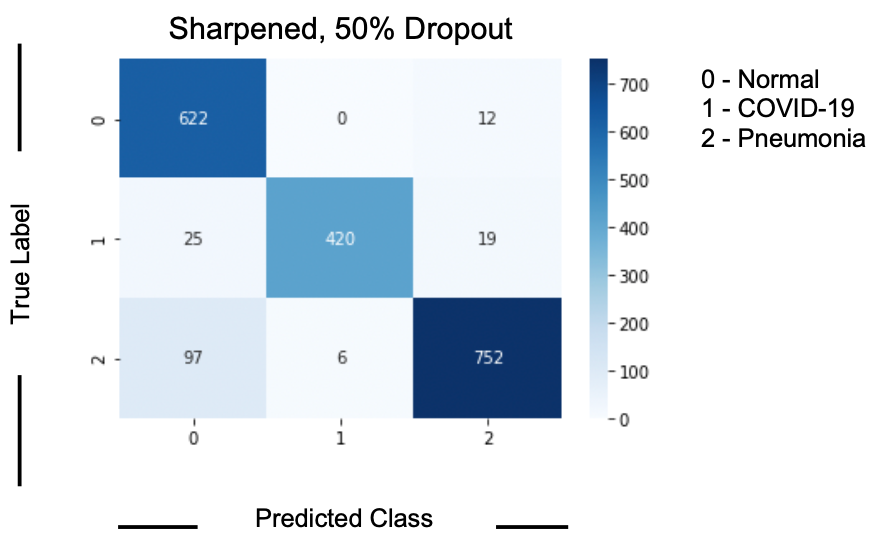}
\end{figure}

\clearpage
\newpage
\section{Appendix D}
\label{sec:D}
\emph{Original Dataset Feature Extraction by Layer of Normal Lung X-Ray}
\begin{figure}[ht!]
\includegraphics[width=\textwidth]{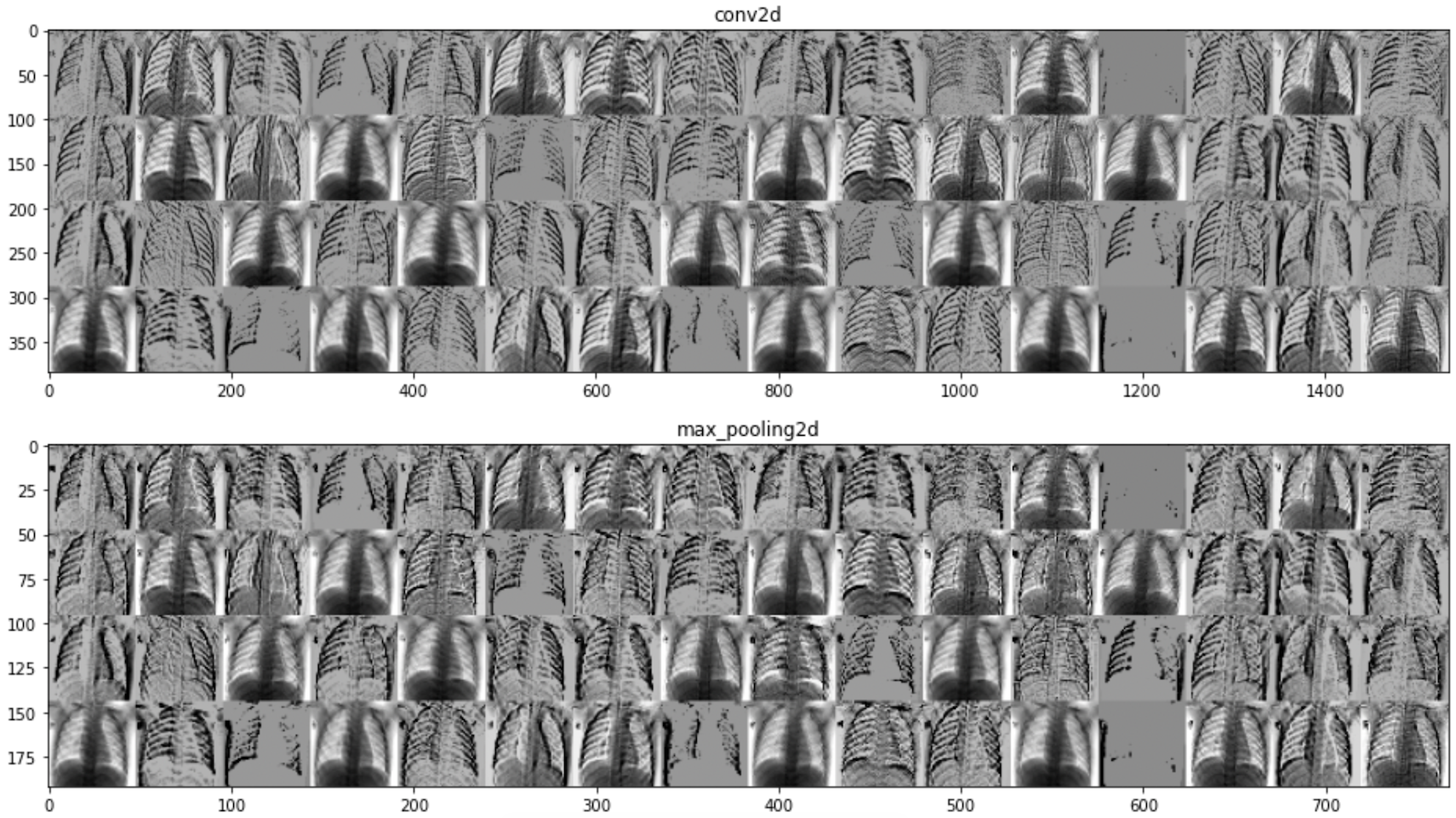}
\end{figure}
\begin{figure}[ht!]
\includegraphics[width=\textwidth]{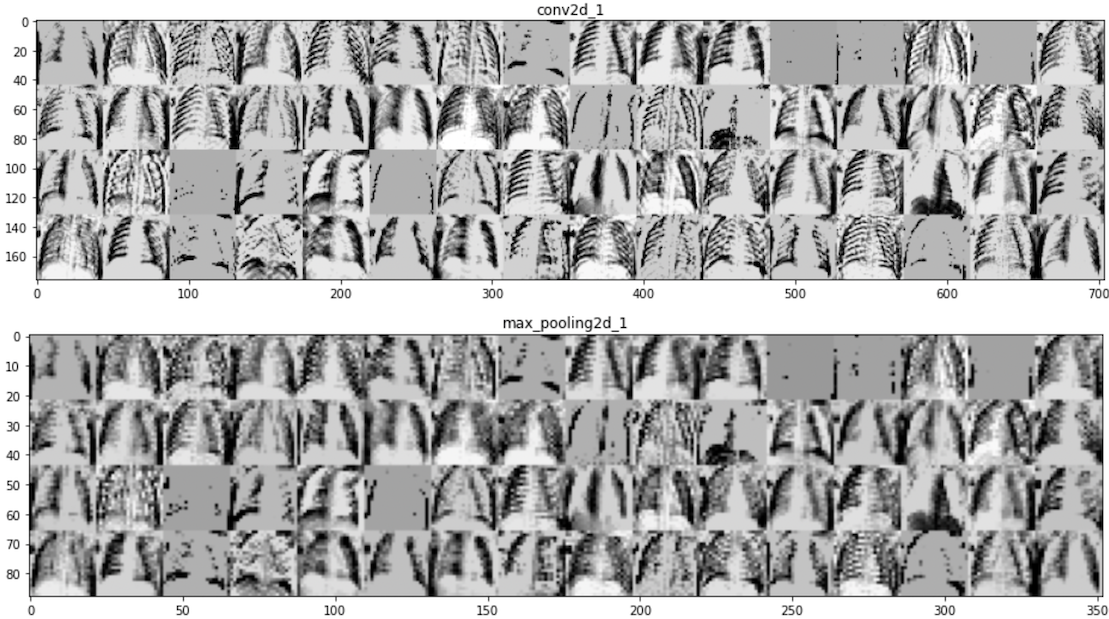}
\end{figure}

\clearpage
\newpage
\section{Appendix E}
\label{sec:E}
\emph{Contoured Dataset Feature Extraction by Layer of Normal Lung X-Ray}
\begin{figure}[ht!]
\includegraphics[width=\textwidth]{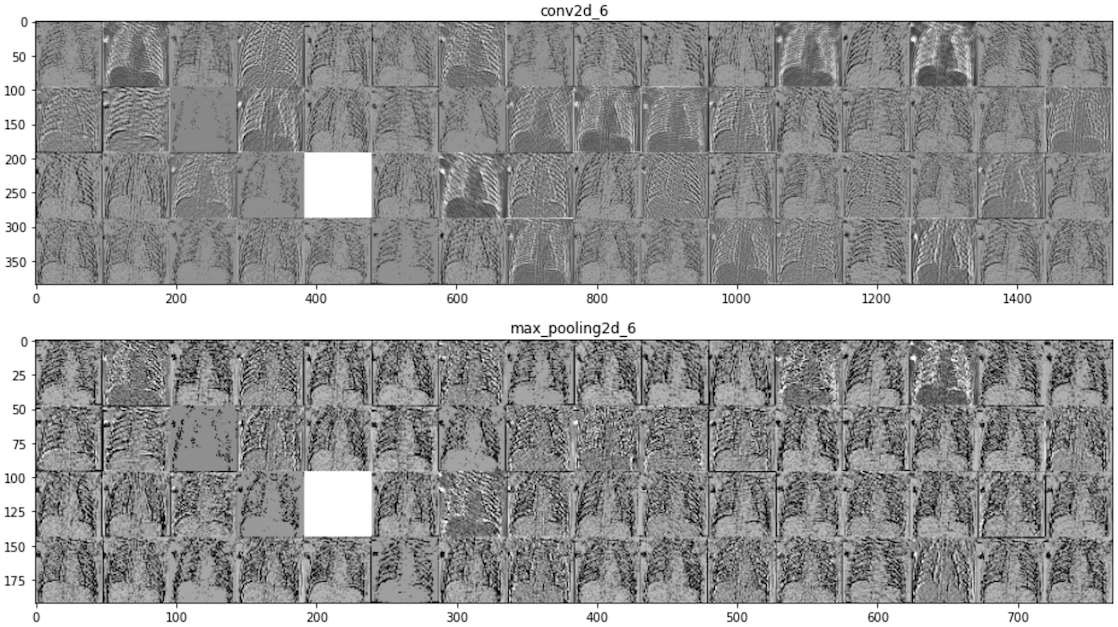}
\end{figure}
\begin{figure}[ht!]
\includegraphics[width=\textwidth]{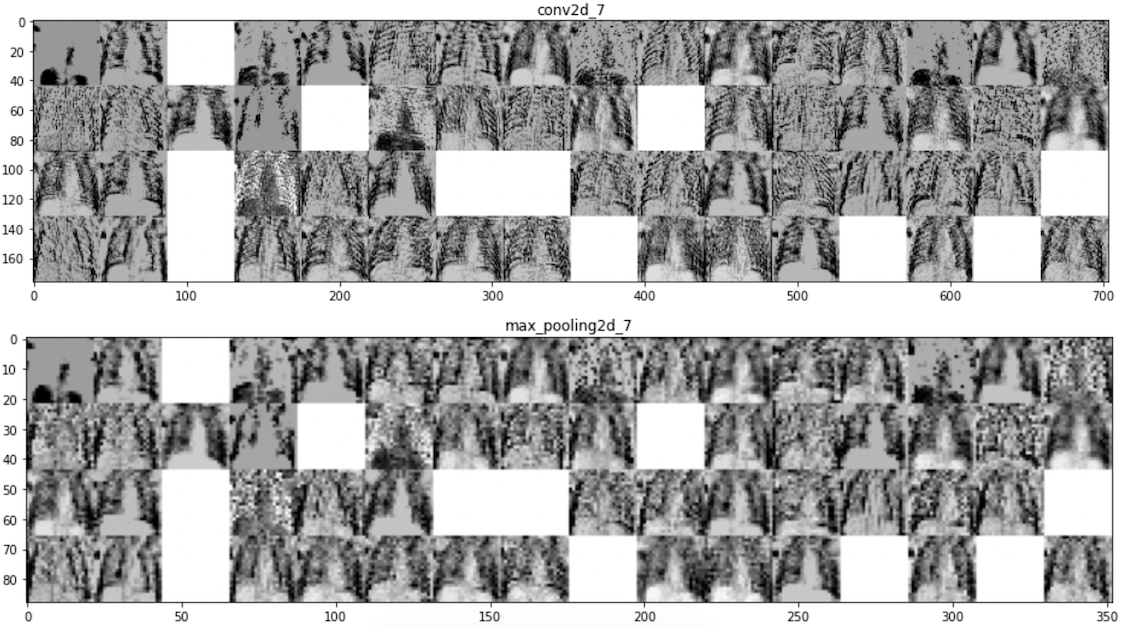}
\end{figure}

\clearpage
\newpage
\section{Appendix F}
\label{sec:F}
\emph{More Edge Enhanced Dataset Feature Extraction by Layer of Normal Lung X-Ray}
\begin{figure}[ht!]
\includegraphics[width=\textwidth]{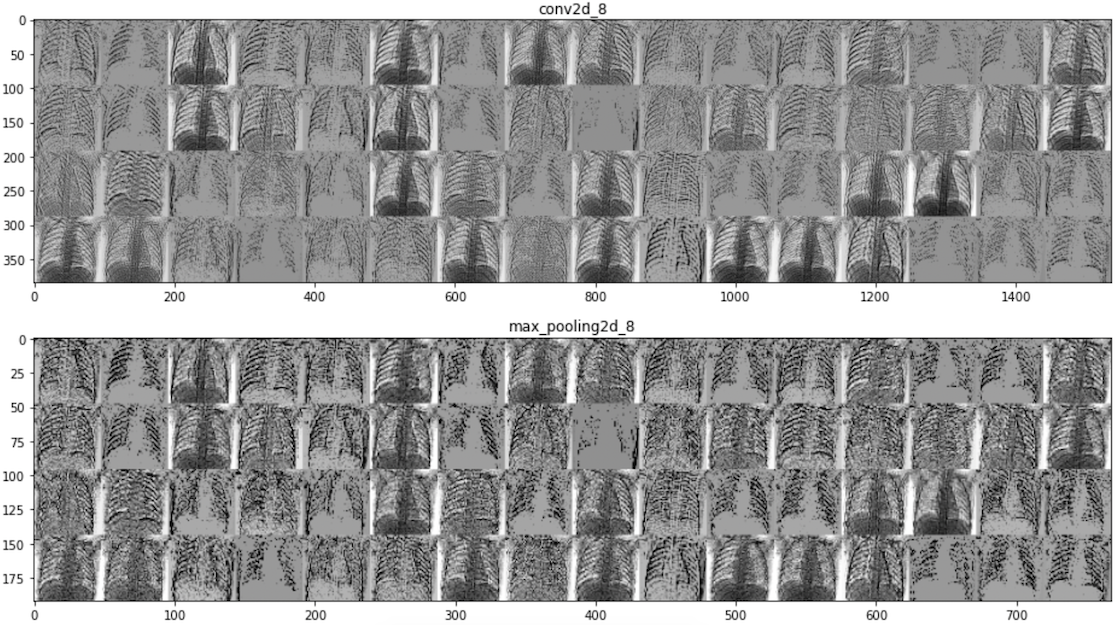}
\end{figure}
\begin{figure}[ht!]
\includegraphics[width=\textwidth]{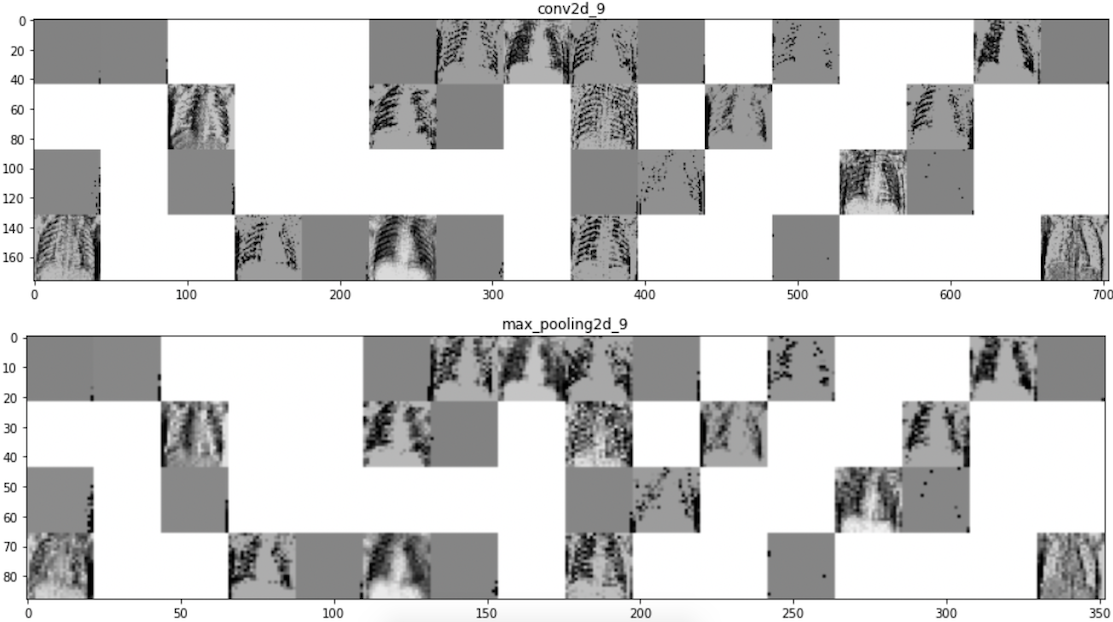}
\end{figure}

\clearpage
\newpage
\section{Appendix G}
\label{sec:G}
\emph{Found Edge Dataset Feature Extraction by Layer of Normal Lung X-Ray}
\begin{figure}[ht!]
\includegraphics[width=\textwidth]{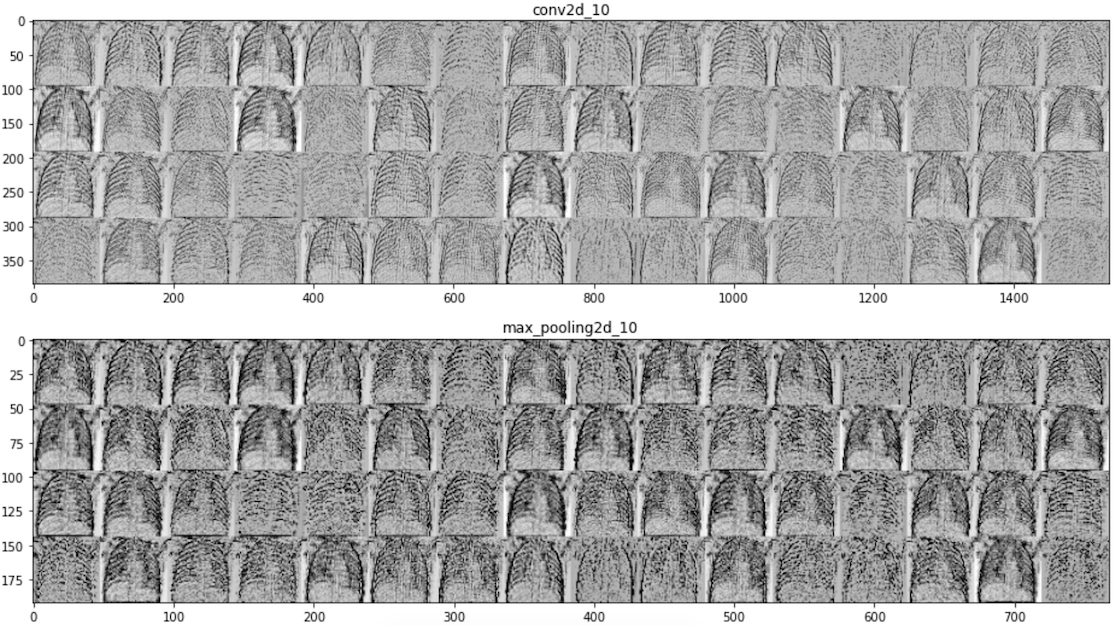}
\end{figure}
\begin{figure}[ht!]
\includegraphics[width=\textwidth]{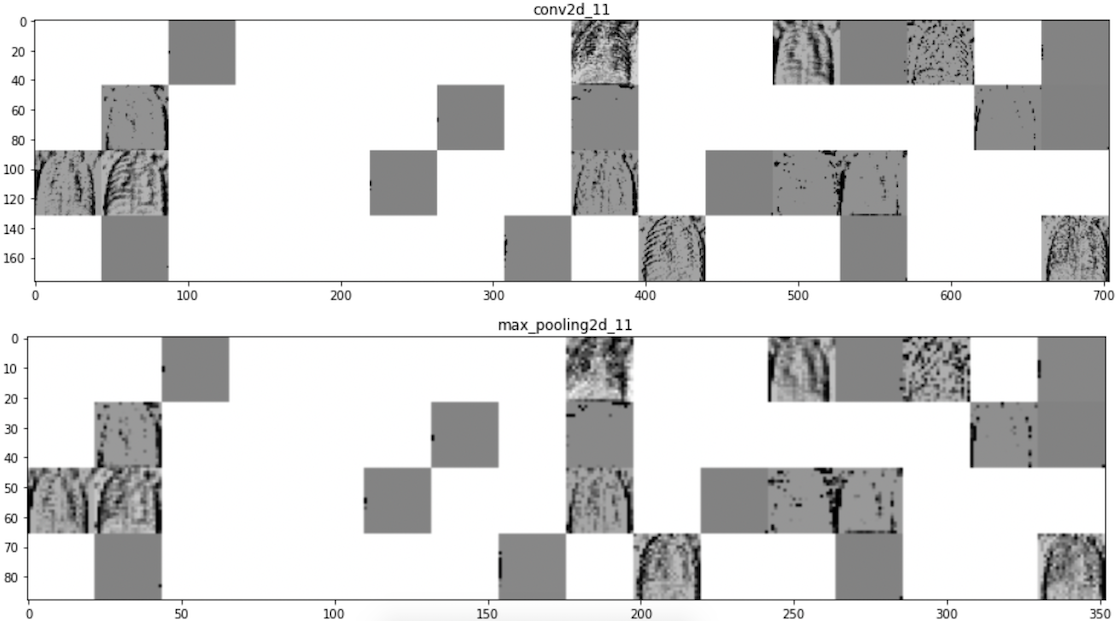}
\end{figure}

\clearpage
\newpage
\section{Appendix H}
\label{sec:H}
\emph{Sharpened Dataset Feature Extraction by Layer of Normal Lung X-Ray}
\begin{figure}[ht!]
\includegraphics[width=\textwidth]{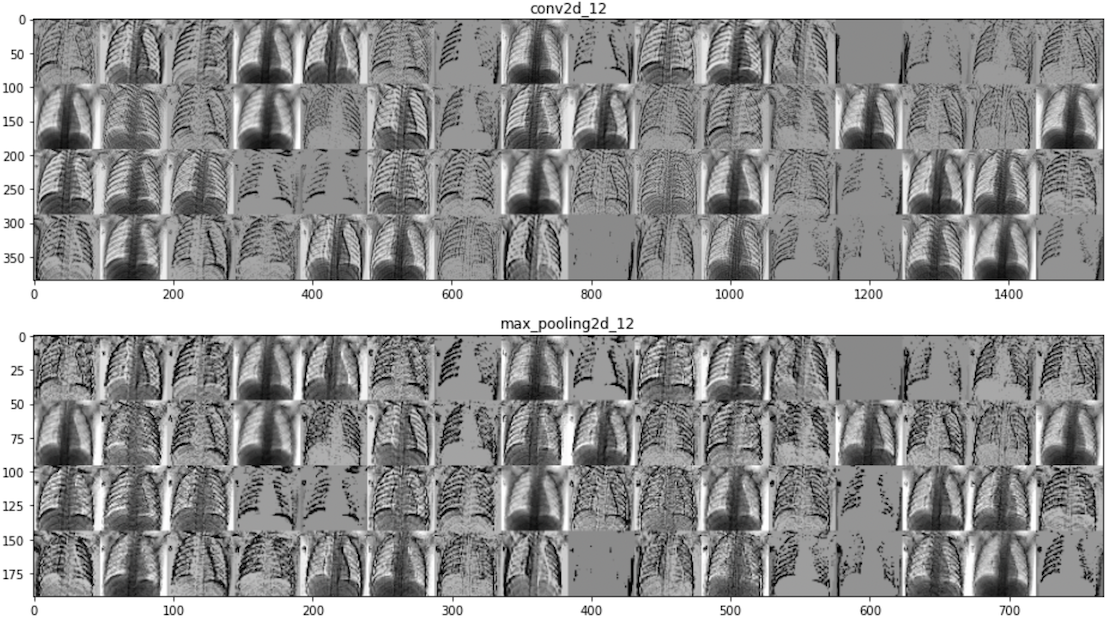}
\end{figure}
\begin{figure}[ht!]
\includegraphics[width=\textwidth]{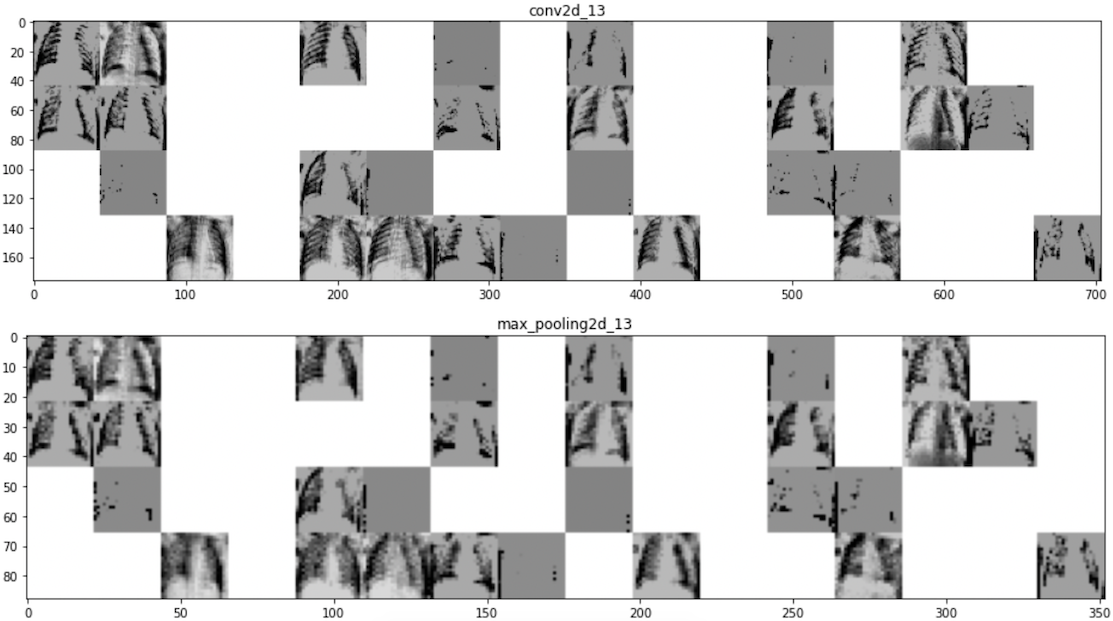}
\end{figure}

\clearpage
\newpage
\section{Appendix I}
\label{sec:I}
\emph{Sharpened, 20\% Dropout Dataset Feature Extraction by Layer of Normal Lung X-Ray}
\begin{figure}[ht!]
\includegraphics[width=\textwidth]{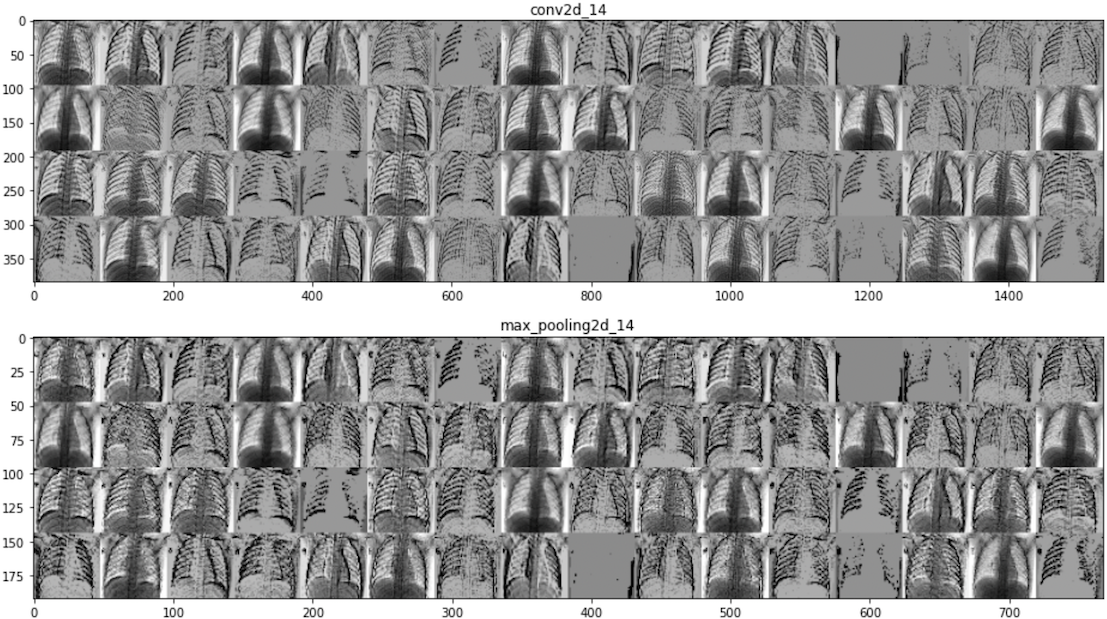}
\end{figure}
\begin{figure}[ht!]
\includegraphics[width=\textwidth]{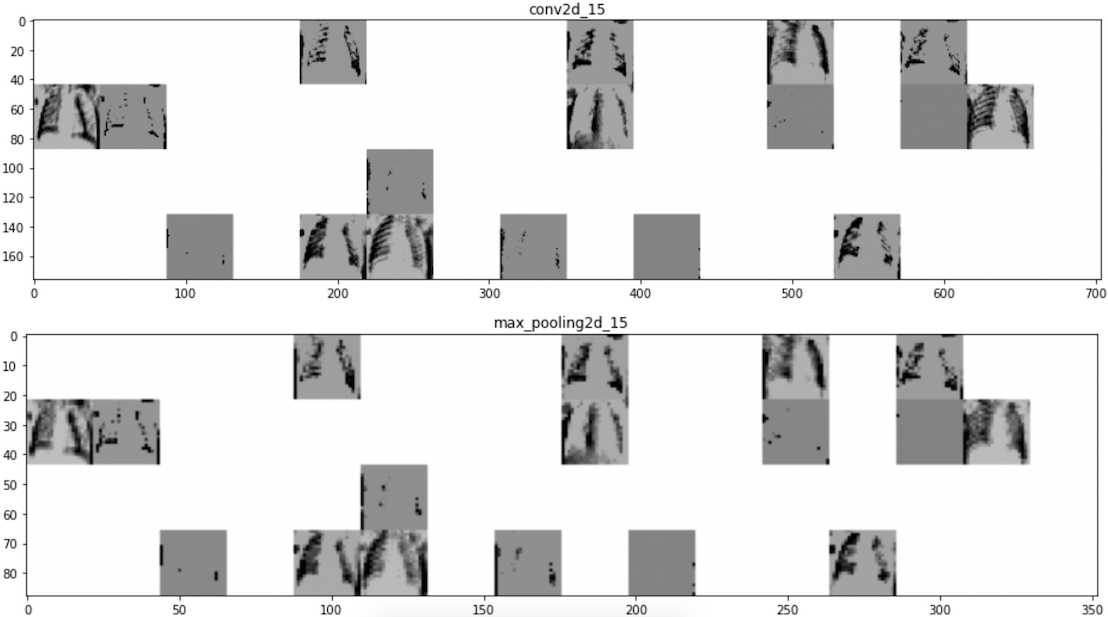}
\end{figure}

\clearpage
\newpage
\section{Appendix J}
\label{sec:J}
\emph{Sharpened, 50\% Dropout Dataset Feature Extraction by Layer of Normal Lung X-Ray}
\begin{figure}[ht!]
\includegraphics[width=\textwidth]{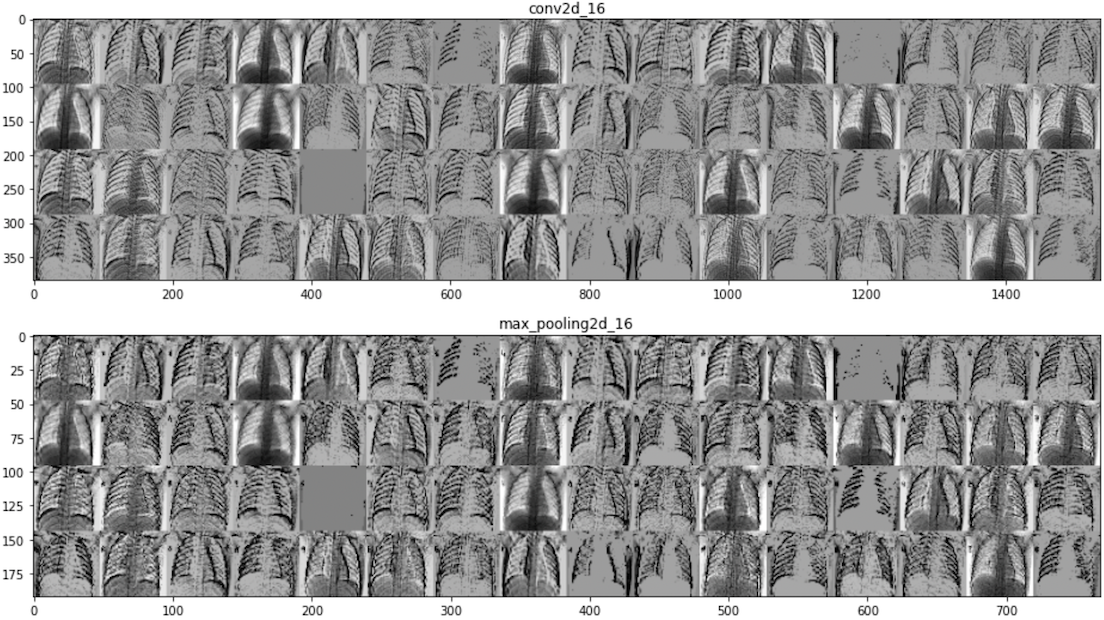}
\end{figure}
\begin{figure}[ht!]
\includegraphics[width=\textwidth]{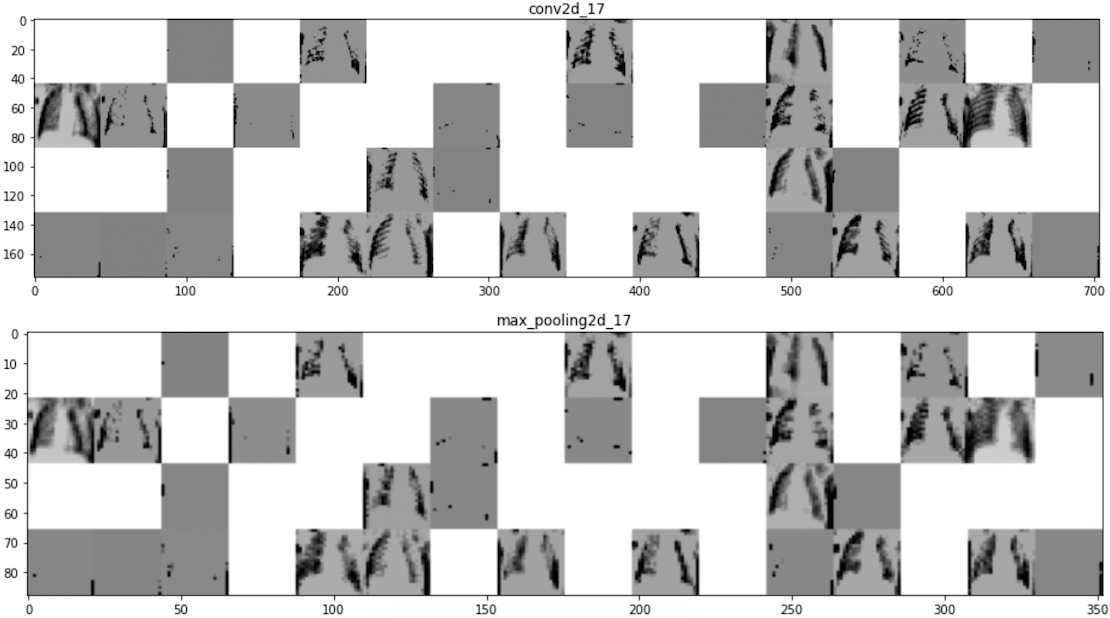}
\end{figure}